\documentclass[aps,nofootinbib,showpacs,preprintnumbers,amsmath,amssymb]{revtex4}

\usepackage{graphicx}
\usepackage{dcolumn}
\usepackage{bm}
\usepackage{color}
\usepackage[pdftex]{epsfig}  

\def\square{\kern1pt\vbox{\hrule height 1.2pt\hbox{\vrule width 1.2pt\hskip 3pt
   \vbox{\vskip 6pt}\hskip 3pt\vrule width 0.6pt}\hrule height 0.6pt}\kern1pt}

\newcommand{\be}{\begin{equation}}
\newcommand{\ee}{\end{equation}}
\newcommand{\bea}{\begin{eqnarray}}
\newcommand{\eea}{\end{eqnarray}}

\newcommand{\nn}{\nonumber}
\newcommand{\vv}[1]{\mathbf{#1}}
\newcommand{\fnl}{f_{\rm NL}}
\newcommand{\gnl}{g_{\rm NL}}

\def\dthreee#1#2{\frac{d^3{#1_{#2}}}{(2\pi)^3}} 

\begin{document}

\title{Non-local bispectra from super cosmic variance}

\author{Bekir Bayta\c{s}}
\email{bub188@psu.edu}
\author{Aruna Kesavan}
\email{aok5232@psu.edu}
\author{Elliot Nelson}
\email{eln121@psu.edu}
\author{Sohyun Park}
\email{spark@gravity.psu.edu}
\author{Sarah Shandera}
\email{shandera@gravity.psu.edu}

\affiliation{Institute for Gravitation and the Cosmos, 
The Pennsylvania State University, University Park, PA 16802, USA}

\begin{abstract}
We present examples of non-Gaussian statistics that can induce bispectra matching local and non-local (including equilateral) templates in biased sub-volumes. We find cases where the biasing from coupling to long wavelength modes affects only the power spectrum, only the bispectrum or both. Our results suggest that ruling out multi-field scenarios is quite difficult: some measurements of the scalar correlations, including the shape of the bispectrum, can be consistent with single-clock predictions even when cosmic variance from super-horizon modes is at work. Furthermore, if observations of the density perturbations rule out single-clock inflation, we will face a serious cosmic variance obstacle in drawing any further conclusions about the particle physics origin of the scalar fluctuations.

\end{abstract}


\preprint{IGC-15/2-1}

\maketitle

\section{Introduction}
\label{sec:intro}

The primordial curvature perturbations are our primary source of information about the inflationary era of the universe. Ideally, we would like to extract an understanding of the particle physics responsible for inflation from the details of the correlations observed in the Cosmic Microwave Background (CMB) inhomogeneities and the large scale structure. However, we observe only a finite volume of the universe and have no reason to expect that the current size of that volume is anything special: there are likely to be at least some super horizon modes with more or less the same properties as those that have already re-entered the horizon. That supposition is critically important in comparing statistical observations to theory. In the case of exactly Gaussian fluctuations the finite size of our Hubble volume is the origin of the familiar cosmic variance uncertainties in the power spectrum fit. 

In non-Gaussian scenarios that couple Fourier modes of very different wavelengths there is additional cosmic variance from the possibility that all our observations are biased compared to the mean predictions of some inflationary theory. The bias from super horizon modes is completely unmeasurable, but the qualitative conclusions we draw about the origin of fluctuations can change when we allow for it \cite{Bartolo:2012sd,Nelson:2012sb,Nurmi:2013xv,LoVerde:2013xka,Bramante:2013moa,LoVerde:2013dgp,Thorsrud:2013mma,Thorsrud:2013kya,Young:2014oea}. Fortunately, no such mode coupling has been detected within our universe yet, but current observational limits (e.g., from the Planck satellite results \cite{Ade:2013ydc} and the most recent constraints from quasars \cite{Leistedt:2014zqa}) do not rule out this possibility. Indeed, many inflation scenarios predict some degree of non-Gaussianity and the details of the correlations would ideally provide a means to distinguish qualitatively different primordial physics. In weighing observational evidence for or against any inflation model that couples modes of different wavelengths we must include these ``Super Cosmic Variance" (SCV) uncertainties. In addition, if {\it any} non-Gaussianity is detected, we must be sure we understand how to draw robust conclusions about the set of primordial universe models consistent with those measurements. This is a non-trivial task because it is not clear if there are types of non-Gaussianity inflation {\it cannot} generate and because it is far from obvious how the statistics in biased sub-volumes are related to those of parent distributions with arbitrary non-Gaussian fluctuations. 

Here we consider several scenarios motivated by predictions from currently studied inflation models and demonstrate how the presence of correlations beyond the bispectrum can alter the shape of lower order correlation functions in biased sub-volumes. We will demonstrate that although observations could prove that the source of the primordial fluctuations was not single-clock inflation, constraints on or detections of the shape of the bispectrum can be consistent with single-clock predictions even when super cosmic variance is at work. 

\subsection{The model}
\label{sec:model}
An attractive discriminating feature of inflation scenarios is the behavior of the squeezed limit of the primordial bispectrum: that is, how significantly it couples two short wavelength modes to one long wavelength mode (represented by a squeezed triangle in momentum space, with side lengths $k_1\equiv k_l\ll k_2\approx k_3\equiv k_s$) \cite{Creminelli:2004yq}. This limit also indicates how significantly the bispectrum can cause the power spectrum in biased sub-volumes to differ from the global power spectrum (for a concrete example entirely within our universe, see the discussion of non-Gaussian halo bias \cite{Dalal2008, Chiang:2014oga}). A bispectrum of the type generated by single clock inflation, which primarily couples modes of the same wavelength, will give a negligible shift to the power spectrum regardless of the realization of the long wavelength modes. A local type coupling, on the other hand, can give an interesting amplitude shift \cite{Nelson:2012sb, LoVerde:2013xka}. 
 
Beyond the bispectrum, all higher order correlation functions can contribute to the biasing of lower order statistics when some modes have longer wavelengths than the size of the spatial region of interest. Scenarios where the non-Gaussian field is any local but non-linear function of a Gaussian field have the nice property that the bispectrum is of the standard local type (up to at most logarithmic corrections) even in biased sub-volumes. And, in sufficiently biased sub-volumes the observed statistics of {\it any} local, non-linear function (polynomial and without derivatives) of a Gaussian field will be those of a field with the local ansatz \cite{Nelson:2012sb}.

Here, we explore the question of biased statistics for scenarios with non-local bispectra. Our goal is to understand how observations of a necessarily limited set of correlation functions can constrain the space of models of the primordial universe. In particular, the single-clock inflation consistency relations \cite{Maldacena:2002vr, Creminelli:2004yq, Cheung:2007sv, Senatore:2012wy, Assassi:2012zq, Goldberger:2013rsa, Hinterbichler:2013dpa,Joyce:2014aqa, Mirbabayi:2014zpa} indicate how one could rule out single clock inflation, but would a detection of $\fnl^{\rm local}=0$ and $\fnl^{\rm equil}\neq0$ confirm single clock inflation? To address this question we consider bispectral shapes characterized by their degree of divergence with the long wavelength mode, $k_l$, in the squeezed limit, including $k_l^{-1}$ (equilateral type), $k_l^{-2}$ (sometimes called `orthogonal' type\footnote{A different, also very useful definition of `orthogonal' comes from looking at possible scale-invariant shapes produced by single clock inflation \cite{Senatore:2009gt}. With that definition, the orthogonal template diverges as $k_l^{-1}$.}), and $k_l^{-3}$ (local type). 

To consider non-Gaussian scenarios with the desired $n$-point correlations, we build the field $\Phi$ from a series of nonlocal functionals of a Gaussian random field $\phi(x)$,
\be\label{ansatz}
\Phi[\phi(\vv{x})]=\phi(\vv{x})+ f_{\rm NL} \Phi_2[\phi(\vv{x})]+g_{\rm NL} \Phi_3[\phi(\vv{x})]+\cdots\;,
\ee
where the subscript on $\Phi_n$ indicates how many copies of the Gaussian field appear in the term. The $\Phi_n$ will generate the connected parts of the tree-level correlations at order $n+1$ and higher, and we require $\langle\Phi_n\rangle=0$. We assume the Gaussian field is homogeneous and isotropic. Its statistics are completely determined by the power spectrum, which we take to be scale-invariant for simplicity:
\bea
\langle\phi(\vv k_1)\phi(\vv k_2)\rangle&=&(2\pi)^3\delta^3(\vv k_1+\vv k_2)P_{\phi}(k_1)\\
P_{\phi}(k)&=&2\pi^2\frac{\Delta^2_{\phi}}{k^3}\;.
\eea

We will primarily work with the Fourier transform of $\Phi$:
\bea\label{kspacePhi}
\Phi(\vv k)&=&\phi(\vv k)+f_{\rm NL}\Phi_2(\vv{k})+g_{\rm NL}\Phi_3(\vv{k})+\dots\nn\\
&=&\phi(\vv k)+\frac{f_{\rm NL}}{2!}\int\dthreee p1\int d^3p_2 \,[\phi(\vv p_1)\phi(\vv p_2)-\langle\phi(\vv p_1)\phi(\vv p_2)\rangle]\,N_2(\vv p_1, \vv p_2,\vv k)\delta^{3}(\vv k-\vv p_1-\vv p_2)\nn\\
&&+\frac{g_{\rm NL}}{3!(2 \pi)^6}\prod_{\ell=1}^3\int  d^3 p_\ell \left[\phi(\vv p_1)\phi(\vv p_2)\phi(\vv p_3)-\sum_{\substack{i=1\\k\neq j\neq i}}^3\phi(\vv p_i)\langle\phi(\vv p_j)\phi(\vv p_k)\rangle\right]N_3(\vv p_1,\vv p_2,\vv p_3,\vv k)\delta^{3}(\vv k-\sum_{\ell=1}^3\vv p_{\ell})\\\nonumber
&&+\cdots\;,
\eea
The kernels $N_n(\vv p_1, \vv p_2,\dots, \vv p_n, \vv k)$ are symmetric in the first $n$ entries. The subtracted terms inside the square brackets maintain $\langle\Phi_n\rangle=0$ and ensure that only connected parts of the $n$-point functions are generated by each term. Loop corrections to the power spectrum from the non-Gaussian terms go like $(f_{\rm NL}\Delta_{\phi})^2$ and $(g_{\rm NL}\Delta_{\phi}^2)^2$ and the one-loop correction to the bispectrum goes like $g_{\rm NL}\Delta_{\phi}^{2}$ (relative to the $\mathcal{O}(f_{\rm NL})$ tree-level bispectrum). Weak non-Gaussianity is defined by requiring those quantities to be $\ll 1$, and if $g_{\rm NL}\lesssim\mathcal{O}(f_{\rm NL}^2)$ the loop correction to the power spectrum from the three point function is parametrically larger than that from the four-point. 

To generate our full set of desired correlations, the $\Phi_n$ must be non-local. It is notationally easier to first generate the appropriate set of terms in real space, so we consider $\Phi_n$ of the form
\bea
\Phi_n = \partial^{\rm \beta_{\rm 2n-1}}(\partial^{\rm \beta_{\rm 2n-2}} \phi ( \dots (\partial^{\rm \beta_2} \phi \partial^{\rm \beta_1} \phi)))(\vv x)\;,
\eea
where the $\beta_i$ can be negative. To generate the corresponding Fourier transformed terms, we define derivative operators acting on $\phi(\vv{x})$ based on the corresponding momentum space behavior,\footnote{Note that $\partial^n$ is not in general a genuine derivative operator because it does not obey the Leibniz rule,
\bea
[\partial^n(\phi_1\phi_2)](\mathbf k)&=&k^n\int\frac{d^3 p}{(2\pi)^3}\phi_1(\mathbf p)\phi_2(\mathbf k - \mathbf p) \nn\\
\left[\phi_1\partial^n\phi_2+\phi_2\partial^n\phi_1\right](\mathbf k)&=&\int\frac{d^3 p}{(2\pi)^3}\phi_1(\mathbf p)\phi_2(\mathbf k - \mathbf p)(p^n+|\mathbf k - \mathbf p|^n) \nn\\
\rightarrow \partial^n(\phi_1\phi_2)&\neq& \phi_1\partial^n\phi_2+\phi_2\partial^n\phi_1.\label{Leibniz}
\eea
Our notation is related to the one used in Scoccimarro et al \cite{Scoccimarro:2011pz} by: $(\partial^2)_{\rm here}=(-\nabla^2)_{\rm there}$.}
\be
\partial^n\phi(\mathbf x)\equiv\int\frac{d^3k}{(2\pi)^3}k^n\phi(\vv k)e^{i\mathbf{k}\cdot\mathbf{x}}\;.
\ee

A $\Phi_2$ that can generate the equilateral, orthogonal or local bispectral templates was derived by Scoccimarro et al in \cite{Scoccimarro:2011pz}. In the next section we review the choice of quadratic terms, taking a slightly different perspective but arriving at the same $\Phi_2$ as \cite{Scoccimarro:2011pz}. In Section $\ref{sec:cubic}$ we generalize the procedure for generating the quadratic terms and consider a $\Phi_3$ designed to induce only the terms in $\Phi_2$ in biased sub-volumes. In Section $\ref{sec:families} $ we examine subsets of the cubic terms that restrict the bispectra induced in biased sub-volumes to local, equilateral, or orthogonal type and consider the implications for theory and observations. In Section $\ref{sec:conclusions}$ we conclude and comment on how our analysis might be further developed.

\section{Quadratic Terms}
\label{sec:quadratic}
In this section we derive an expression for $\Phi_2$ using a procedure that can be easily generalized to the cubic term, $\Phi_3$, and beyond. Although our motivation and procedure are slightly different, the result reproduces the expressions derived by Scoccimarro et al \cite{Scoccimarro:2011pz}. 

We limit our considerations to bispectra that diverge in the squeezed limit as one, two, or three inverse powers of the long wavelength momentum. Since a local quadratic function, $\phi^2(x)$, yields the most divergent bispectrum, we need to include inverse derivative operators in order to reduce the divergence. We therefore start with the following family of terms quadratic in the Gaussian field
\be
\partial^{\alpha_3}(\partial^{\alpha_2}\phi\partial^{\alpha_1}\phi),
\label{quad_form}
\ee 
with restrictions on the $\alpha_i$:
\begin{itemize}
\item $\sum\alpha_i=0$. This condition maintains scale-invariance.
\item $\alpha_{1,2}\geq 0$. Together with the previous condition, this automatically sets $\alpha_3 \leq 0$. This condition ensures that the infrared (IR) sensitivity to $\phi_l$ is no stronger than that of local non-Gaussianity. In other words, this rule ensures that the squeezed limit of the bispectrum does not grow like $k_l^{-4}$ or more for small $k_l$.

\item $|\alpha_i|\leq 2$. This restriction generates the {\it minimal} set of terms required to produce equilateral ($k_l^{-1}$), orthogonal ($k_l^{-2}$) or local ($k_l^{-3}$) type behavior (and corresponds to to setting $u=s=0$ in \cite{Scoccimarro:2011pz}). However, there are certainly additional terms with $|\alpha_i|> 2$ that will also generate bispectra with the same squeezed limit behavior. We comment on relaxing this condition in Section \ref{sec:conclusions}. 

\end{itemize}
The generic quadratic functional with terms obeying these constraints is
\be
\Phi_2[\phi(x)] = [ a_1\phi^2+a_2\partial^{-1}(\phi\partial\phi)+a_3\partial^{-2}(\phi\partial^2\phi)+a_4\partial^{-2}(\partial\phi)^2] - {\rm [E.V.]} \;.
\label{Phi_quad}
\ee
where $-[{\rm E.V.}]$ indicates that the expectation values of the terms should be subtracted. The corresponding $N_2$ kernel (which in this case depends only on the magnitudes of the momenta) is
\be
N_2(p_1,p_2,k) = 2a_1+a_2\frac{p_1+p_2}{k}+a_3\frac{p_1^2+p_2^2}{k^2}+2a_4\frac{p_1 p_2}{k^2}\;
\label{N_2}\;.
\ee
A generic homogeneous and isotropic bispectrum for the potential $\Phi$ can be written as
\be
\label{eq:threepoint}
\langle\Phi(\mathbf k_1)\Phi(\mathbf k_2)\Phi(\mathbf k_3)\rangle=(2\pi)^3\delta^3(\mathbf k_1+\mathbf k_2+\mathbf k_3)\;B(k_1,k_2,k_3)\;.
\ee
Our ansatz gives a bispectrum
\be
B_{\Phi}(k_1,k_2,k_3)=f_{\rm NL}P_{\phi}(k_1)P_{\phi}(k_2)N_2(k_1,k_2,k_3)+\mbox{cyc.}\;
\label{bispectrum}
\ee
where there are 2 additional cyclic permutations. Notice that both the $a_2$ and $a_3$ terms in Eq.(\ref{N_2}) generate terms of the type $P(k_1)^{1/3}P(k_2)^{2/3}P(k_3)$ in the bispectrum.

We have already factored out an overall amplitude, $f_{\rm NL}$, from $\Phi_2$. With the usual convention $f_{\rm NL}\equiv {6}B_{\Phi}(k,k,k)/P^2_{\Phi}(k)$, this leaves a normalization condition for the coefficients of the individual quadratic terms: 
\begin{equation}
a_{1} + a_{2} + a_{3} + a_{4} = 1\;.
\label{normalizea}
\end{equation}

The constraints on the $\alpha_i$ so far allow us to restrict consideration to a subset of possible non-Gaussian fields based on the behavior of the bispectrum. However, any quadratic term will also contribute to the power spectrum of the full, non-Gaussian field
\be
\langle\Phi(\mathbf k_1)\Phi(\mathbf k_2)\rangle=\langle\phi(\mathbf k_1)\phi(\mathbf k_2)\rangle +f_{\rm NL}^2\langle\Phi_2(\mathbf k_1)\Phi_2(\mathbf k_2)\rangle+\dots
\ee
where the dots contain contributions from $\Phi_3$ and higher. The contributions to $P_{\Phi}(k)$ from the terms in $\Phi_2$ contain an extra integral over momenta $p$ and go like $[N_2(p,|\vv k-\vv p|,k)]^2$. The contributions from the $a_3$ and $a_4$ terms go like $1/k^4$ times a divergent integral over momenta $p$ (this is easiest to see by considering $k\ll p$). We can force this badly behaved contribution to the loop to vanish by setting
\be
a_4=-a_3\;.
\label{1loopconstraint}
\ee
We will impose this condition from now on to remove $a_4$ from all expressions. Insisting on Eq.(\ref{1loopconstraint}) is equivalent to the choice of the coefficient $t$ for the equilateral and orthogonal cases in \cite{Scoccimarro:2011pz}.

\subsection{Recovering the standard bispectral templates}

Frequently used bispectral templates with fixed degree of divergence with the long wavelength mode are
\bea
\label{templates}
B_{\rm local}&=&2f_{\rm NL}^{\rm local}(P_{\phi}(k_1)P_{\phi}(k_2) +P_{\phi}(k_1)P_{\phi}(k_3) +P_{\phi}(k_2)P_{\phi}(k_3))\\\nonumber
B_{\rm equil}&=&6f_{\rm NL}^{\rm equil}[-P_{\phi}(k_1)P_{\phi}(k_2) +{\rm 2\;perm.} -2(P_{\phi}(k_1)P_{\phi}(k_2)P_{\phi}(k_3))^{2/3}\\\nonumber
&&+P_{\phi}(k_1)^{1/3}P_{\phi}(k_2)^{2/3}P_{\phi}(k_3)+{\rm 5\;perm.} ]\\\nonumber
B_{\rm orth}&=&6f_{\rm NL}^{\rm orth}[- 3P_{\phi}(k_1)P_{\phi}(k_2) +{\rm 2\;perm.}-8(P_{\phi}(k_1)P_{\phi}(k_2)P_{\phi}(k_3))^{2/3}\\\nonumber
&&+3P_{\phi}(k_1)^{1/3}P_{\phi}(k_2)^{2/3}P_{\phi}(k_3) +{\rm 5\;perm.} ]
\eea

The best constraints on the amplitudes $f_{\rm NL}$ of these templates come from the {\it Planck} satellite:\cite{Ade:2013ydc}, which limit $f_{\rm NL}^{\rm local}=2.7\pm5.8$, $f_{\rm NL}^{\rm equil}=-42\pm75$, and $f_{\rm NL}^{\rm orth}=-25\pm39$ at the 68.3\% confidence level.

The ansatz in Eq.(\ref{Phi_quad}) clearly contains the local ansatz
\be
{\rm Local\; Bispectrum:}\;\;\;\;\;\;a_2=a_3=a_4=0\;.
\ee
 
 To see the conditions that recover the orthogonal and equilateral templates, insert the $N_2$ kernel from Eq. (\ref{N_2}) into the general expression for the bispectrum, Eq. (\ref{bispectrum}), and take the squeezed limit $k_1\equiv k_l\ll k_2,k_3\equiv k_s$:
\bea \label{squeezed_bi}
\lim_{k_l\ll k_s}B(k_s,k_s,k_l) = f_{NL} P_{\phi}(k_l)P_{\phi}(k_s)
\left[(4a_1+2a_2+2a_3)+(2a_2-4a_3)\Big(\frac{k_l}{k_s}\Big)+(2a_2+2a_3)\Big(\frac{k_l}{k_s}\Big)^2+2a_1\Big(\frac{k_l}{k_s}\Big)^3\right].
\eea
where we have already used $a_4=-a_3$. The equilateral bispectrum scaling $k_l^{-1}$ requires that the coefficients of both the $k_l^{-3}$ and the $k_l^{-2}$ contributions vanish. Since $P(k_l)\propto k_l^{-3}$, these conditions are, respectively

\be
4a_{1}+2a_{2}+2a_{3}= 0 \;\;\;\;\;\; {\rm and} \;\;\;\;\;\;2a_2-4a_3=0\;.
\label{Bsq_kl^-2-3zero}
\ee
Including the normalization condition \eqref{normalizea}, these are sufficient equations to uniquely fix the $a_i$. The result exactly recovers the equilateral template:
\begin{equation} \label{equil}
{\rm Equilateral\; Bispectrum:}\;\;\;\;\;\;a_{1} = -3, \, a_{2} = 4, \, a_{3} = 2, a_{4} = -2\;.
\end{equation}

The orthogonal bispectrum scaling $k_l^{-2}$ requires only the coefficient of $k_l^{-3}$ to vanish, which is not sufficient information to completely fix all the $a_i$. That is reasonable since the orthogonal shape was originally defined not by its squeezed limit but by minimizing its overlap with the other templates over a range of momentum configurations. To generate the standard orthogonal template we can add an additional condition, comparable to the orthogonality condition required by \cite{Senatore:2009gt}, 
\be\label{exconstr}
a_{1}^{\rm orth}= 3 \,\, a_{1}^{\rm equil}
\ee
so that all the $a_i$ are fixed:
\be
{\rm Orthogonal \;Bispectrum:}\;\;\;\;\;\;a_{1} = -9, \, a_{2} = 10, \, a_{3} = 8, a_{4} = -8\;.
\ee 

In the next section we will introduce the long-short wavelength split and see how the conditions in Eq.(\ref{Bsq_kl^-2-3zero}) that fix the squeezed limit behavior of the bispectrum can also be directly read off of the real-space expression in Eq.(\ref{Phi_quad}).

\subsection{The long-short wavelength split}

We are interested in the effect of long wavelength background modes on the statistics measured in spatial sub-volumes. To that end, we split the field $\rm \Phi$ in Fourier space at a scale $k_*$,
\bea \label{split}
\Phi_{s}(\vv k) = \Phi(\vv k)\Theta(k-k_*) \nn \\
\Phi_{l}(\vv k) = \Phi(\vv k)\Theta(k_*-k)\;,
\eea
where $\Theta$ is the step function. This is only an approximation to splitting the field with a top hat in real space, but our results will not depend on this distinction. Also, in what follows, we will mainly consider momenta far away from the scale $k_*$, and ignore complications arising from those close to $k_*$.
Applying the $\Theta$-function on the right hand side of $\Phi(\vv{k})$, defined in Eq.\eqref{kspacePhi}, gives
\bea
\Phi_{s}(\vv k)=\phi_{s}(\vv k) + f_{\rm NL} \Phi_{2}^{(s)}(\vv k)+\dots\\
\Phi_{l}(\vv k)=\phi_{l}(\vv k) + f_{\rm NL} \Phi_{2}^{(l)}(\vv k)+\dots
\eea
where $\phi_s(\vv k)$ and $\phi_l(\vv k)$ are defined analogously to Eq.\eqref{split}. Writing out the effect of the $\Theta$-function on the quadratic term gives, for the short wavelength quadratic piece,
\bea
\label{eq:shortPhi}
\Phi_{2}^{(s)}(\vv k) &=& \frac{1}{2!(2 \pi)^{3}} \int d^3p_1\, d^3p_2 \,N_2(\vv p_1,\vv p_2, \vv k) [\phi(\vv p_1)\phi(\vv p_2)-{\rm [E.V.]}]\, \delta^3(\vv k - \vv p_1 -\vv p_2)\, 
\Theta(|\vv p_1 + \vv p_2| - k_*) \nn\\
& = &\frac{1}{2!(2 \pi)^{3}} \!\! \int\limits_{\vv |\vv p_1 + \vv p_2| > \vv k_{*}} \!\!\!\!\!\!\!  d^3p_1 d^3p_2 \,N_2(\vv p_1,\vv p_2, \vv k) [\phi_s(\vv p_1)\phi_s(\vv p_2)+2\phi_s(\vv p_1)\phi_l(\vv p_2)-{\rm [E.V.]}]\, \delta^3(\vv k - \vv p_1 -\vv p_2)
\eea
Here, the second line is obtained from by considering the separate momentum regimes for $p_1$ and $p_2$ that satisfy the step function: $\Theta(|\vv p_1 + \vv p_2| - k_*) = [\Theta(p_1 - k_*) \Theta(p_2 - k_*) + \Theta(p_1 - k_*) \Theta(k_* - p_2) + \Theta(k_* - p_1) \Theta(p_2 - k_*)]\Theta(|\vv p_1 + \vv p_2| - k_*)$.
In the same manner we can write $\Phi_2^{(l)}$ (with some redundancy) as
\bea
\label{longsplit}
\Phi_{2}^{(l)}(\vv k) = \frac{1}{2!(2 \pi)^{3}} \!\! \int\limits_{\vv |\vv p_1 + \vv p_2| < \vv k_{*}} \!\!\!\!\!\!\!  d^3p_1\, d^3p_2 \,N_2(\vv p_1,\vv p_2, \vv k) [\phi_s(\vv p_1)\phi_s(\vv p_2)+ \phi_l(\vv p_1)\phi_l(\vv p_2)-{\rm [E.V.]}]\, \delta^3(\vv k - \vv p_1 -\vv p_2)\eea
Notice that in the first term of Eq.\eqref{longsplit}, the angle between short scale modes $\vv p_1$ and $\vv p_2$ must be nearly $\pi$ so they sum to a long mode with magnitude $k$ below the splitting scale. 

\subsection{The field observed in biased sub-volumes}

We now examine the difference in the global and local statistics of the power spectrum when the quadratic terms above are included. Very generally, the observed field, restricted to points $\vv x$ within a sub-volume $V_S$, will take the form
\be
\label{eq:generalObserved}
\Phi^{\rm obs}(\vv x)|_{x\in V_S}=(\phi+\Phi_1^{\rm SCV})+\fnl(\Phi_2+\Phi_2^{\rm SCV})+\gnl(\Phi_3+\Phi_3^{\rm SCV})\dots \;,
\ee
where the contributions from the Super Cosmic Variance (SCV) terms, $\Phi_n^{\rm SCV}$, vanish in sub-volumes where all long wavelength modes take their mean value (0). Since an observer with access to only a single sub-volume cannot separate the mean contributions from the SCV contributions it is more natural to write the right hand side of the equation above in terms of the observer's linear field, $\chi(\vv x)$, and kernels, and amplitudes:
\be
\Phi^{\rm obs}[\chi(\vv x)]|_{x\in V_S}=\chi(\vv x)+\tilde{f}_{\rm NL}\tilde\Phi_2[\chi(\vv x)]+\tilde{g}_{\rm NL}\tilde\Phi_3[\chi(\vv x)]\dots \;,
\ee
Note that the standard way of defining the normalizations $\tilde{f}_{\rm NL}$ and $\tilde{g}_{\rm NL}$ only works for scale-invariant contributions from the SCV terms.

For the set of quadratic terms we are considering here, using the kernel $N_2$ from Eq.\eqref{N_2} in the expression for the short wavelength field, Eq.\eqref{eq:shortPhi} gives
\begin{align}
\Phi_{s} (\vv k) & \approx  \phi_s(\vv k)\bigg\{1 + \fnl\Big[
(2a_1+a_2+a_3)\phi_l+(a_2-2a_3)\partial\phi_l\frac{1}{k}+a_3\partial^2\phi_l\frac{1}{k^2}\Big]\bigg\}\nn \\
& +\frac{f_{\rm NL}}{2!(2\pi)^3}\int\limits_{p_1,p_2>k_*} d^3p_1d^3p_2 N_2(p_1,p_2,k)\bigg[\phi_s(\vv p_1)\phi_s(\vv p_2) - {\rm [E.V.]}\bigg]\delta^{3}(\vv k-\vv p_1-\vv p_2)
\label{Phi_s}
\end{align}
where the approximation comes because we have taken $\vv k - \vv p_2 \simeq \vv k$ in the $\phi_s(\vv p_1)\phi_l(\vv p_2)$ term in Eq.\eqref{eq:shortPhi}. The functions of the long wavelength modes are
\be
\label{eq:phidelphis}
\phi_l \equiv \int_{\Lambda}^{k_*}\frac{d^3p}{(2\pi)^3}\,\phi(\vv p),  \quad   
\partial\phi_l \equiv \int_{\Lambda}^{k_*}\frac{d^3p}{(2\pi)^3}\,p\phi(\vv p), \quad 
\partial^2\phi_l \equiv \int_{\Lambda}^{k_*}\frac{d^3p}{(2\pi)^3}\,p^2\phi(\vv p)
\ee
where $\Lambda$ is an infrared scale corresponding to the longest wave mode that exited the horizon during inflation. We will treat the quantities in Eq.(\ref{eq:phidelphis}) as constants in any particular sub-volume. While this is not exactly true, since a nonzero gradient $\partial\phi_l$ implies non-constant $\phi_l(\mathbf x)$, we show below that this difference is small compared to the effect of the average value of $\phi_l$.

The observed linear term in a biased sub-volume is shifted from the original $\phi(k)$ by a term whose amplitude and scale-dependence depends on the bispectrum and the bias:
\bea
\label{eq:PhiLinQuad}
\chi_{G}(\vv k) &\equiv&\phi_s(\vv k)[1+\fnl \Phi_1^{\rm SCV} (k)]\nn\\
&=&  \phi_s(\vv k)\bigg\{1 + \fnl\Big[
(2a_1+a_2+a_3)\phi_l+(a_2-2a_3)\partial\phi_l\frac{1}{k}+a_3\partial^2\phi_l\frac{1}{k^2}\Big]\bigg\}
\;.
\eea
Then, using the observed linear field everywhere, the observed non-Gaussian field is
\bea
\Phi^{\rm obs} (k)&=& \chi_{G}(\vv k)\\
&& + \frac{\fnl}{2!(2\pi)^3}\int d^3p_1\,d^3p_2 \,\frac{N_2(\vv p_1, \vv p_2,\vv k)}{[1 + \fnl \Phi_1^{\rm SCV}(p_1)] [1 + \fnl \Phi_1^{\rm SCV}(p_2)]}[\chi_{G}(\vv p_1)\chi_{G}(\vv p_2) - {\rm [E.V.]}] \delta^{3}(\vv k-\vv p_1-\vv p_2)\;.\nn
\eea
The power spectrum and bispectrum observed in sub-volumes can be computed as usual from this expression, and will differ in amplitude and scale-dependence (shape) from the corresponding quantities in unbiased sub-volumes. 

The observed power spectrum is shifted from the input power spectrum by (assuming weak non-Gaussianity)
\bea
P_{\Phi}^{\rm obs}\approx P_{\chi}^{\rm obs}(k) = P_{\phi}(k)\bigg\{1 + \fnl\Big[
(2a_1+a_2+a_3)\phi_l+(a_2-2a_3)\partial\phi_l\frac{1}{k}+a_3\partial^2\phi_l\frac{1}{k^2}\Big]\bigg\}^2\;.
\label{P_s}
\eea
This effect was discussed and quantified in detail for the case of local non-Gaussianity in \cite{LoVerde:2013xka}. There it was shown that the correction proportional to $\phi_l$ can be large and interesting. The other terms here, proportional to $\partial \phi_l$ and $\partial^{2} \phi_l$ are sensitive to only a very small range of modes beyond the horizon.  If the universe was inflating for $N_e$ e-folds before the mode of scale $k_*$ exited the horizon, the super cosmic variance contributions from non-local bispectra are of order 
\bea \label{eq:delphis}
\frac{\langle (\partial \phi_l)^{2}\rangle}{k_{*}^{2}}  &=& \frac{1}{k_{*}^{2}} \int_{\Lambda}^{k_*}\!\! dk\, k\, \Delta^2_{\phi}(k) 
= \frac{\Delta^2_{\phi}(k_*)}{(n_s + 1)} \bigg(1-e^{-(n_s+1)N_e}\bigg)\;,\nn\\
\frac{\langle (\partial^2 \phi_l)^{2} \rangle}{k_{*}^{4}} &=& \frac{1}{k_{*}^{4}} \int_{\Lambda}^{k_*}\!\! dk\, k^3\, \Delta^2_{\phi}(k) 
= \frac{\Delta^2_{\phi}(k_*)}{(n_s + 3)} \bigg(1-e^{-(n_s+3)N_e}\bigg)
\eea
Here we have assumed for simplicity that the super horizon power spectrum is $\Delta^2_{\phi}=A_0(k/k_*)^{n_s-1}$ with constant spectral index $n_s$ and used $N_e=ln(k_*/\Lambda)$. Notice that with $k_*=H_0$ the second quantity in Eq.\eqref{eq:delphis} is of the order of the contribution of fluctuations to the spatial curvature, $\langle\Omega_k^2\rangle$. 

\begin{figure}[htbp]
 \centering
 \includegraphics[scale=0.5]{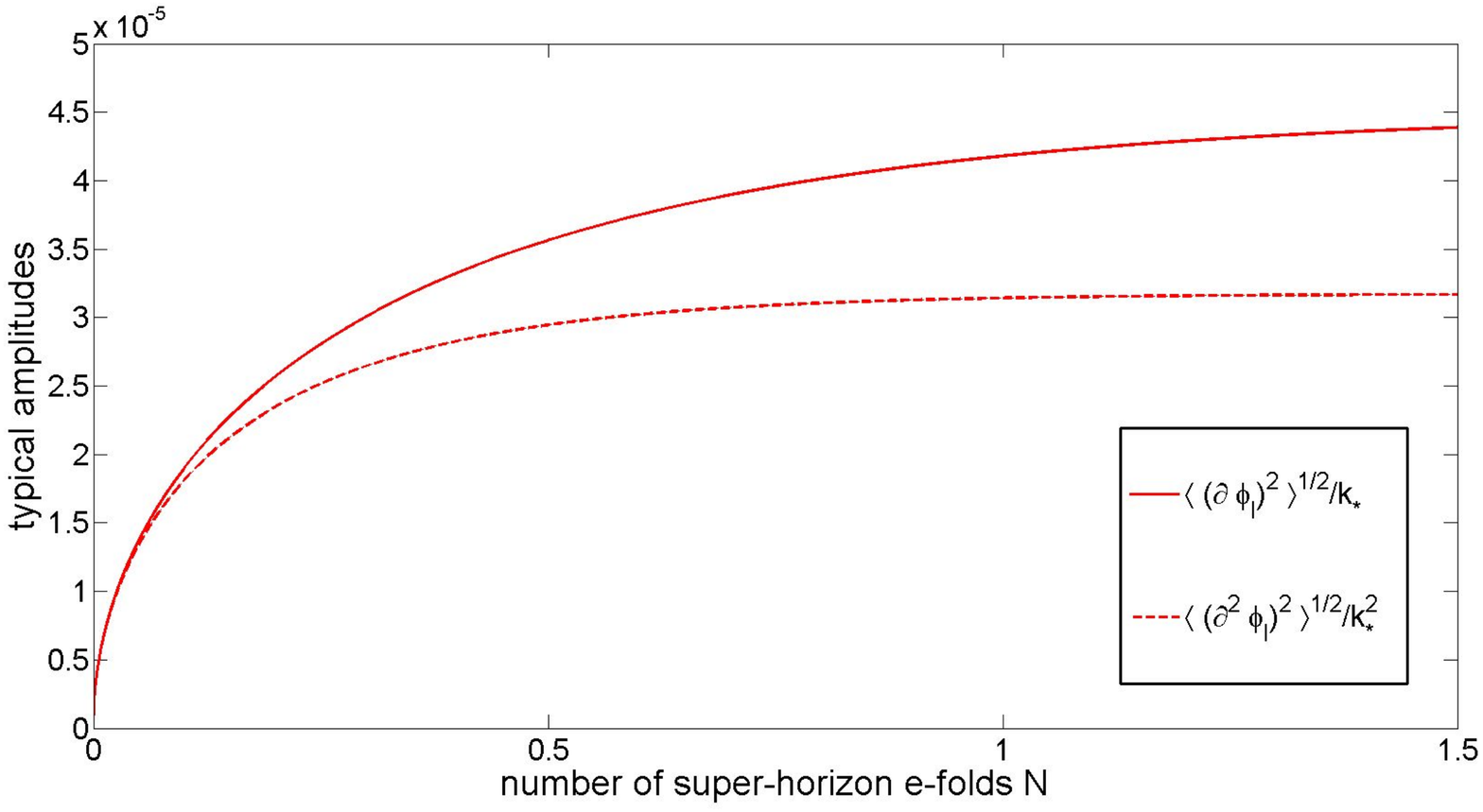}
 \caption{Graph of the RMS amplitude of long-wavelength super cosmic variance terms from non-local bispectra, $\frac{1}{k_*}\langle (\partial \phi_l)^{2} \rangle^{1/2}$ and $\frac{1}{k_*^2}\langle (\partial^{2} \phi_l)^{2} \rangle^{1/2} $. This plot assumes that the power spectrum on scales larger than $k_{*}=0.05 \,\mathrm{Mpc}^{-1}$ remains consistent with {\it Planck} satellite observations: $\Delta^2_{\Phi}(k)=3.98772\times10^{-9}(k/k_{*})^{n_s-1}$ with $n_s=0.9619$ \cite{Ade:2013zuv}. Notice that the plotted quantities are not sensitive to more than about an e-fold of inflation and remain $\mathcal{O}(10^{-5})$ so, as expected, only the local bispectrum can generate a significant super cosmic variance contribution to the power spectrum.}
 \label{figure:typ_orth_equil}
\end{figure}

Figure \ref{figure:typ_orth_equil} shows that the quantities in Eq.(\ref{eq:delphis}) are very small, of order $10^{-5}$. As expected, there is no appreciable cosmic variance shift to the power spectrum, Eq.(\ref{P_s}), from non-local bispectra. Figure \ref{figure:typ_orth_equil} assumes that the power spectrum is consistent with the current best fit from the Planck satellite (Planck+WP) \cite{Ade:2013zuv} ($\Delta^2_{\Phi}(k)=3.98772\times10^{-9}(k/k_{*})^{n_s-1}$, $n_s=0.9619$) and uses $k_{*}=0.05 \,\mathrm{Mpc}^{-1}$ as the reference infrared scale. ($H_0$ is the appropriate scale for considering our current universe.) The strong suppression with extra powers of $k$ in the integrands makes these quantities insensitive to small variations in the spectral index. Specifically, there is no considerable difference between flat spectral index ($n_s=1$) and $n_s=0.9619$ (in contrast to the significant enhancement $\langle\phi_l^2\rangle^{1/2}$ receives from a red tilt over sufficiently many super horizon e-folds). 

Although these terms have a substantial scale-dependence, the maximum shift to the spectral index at the CMB pivot point is of order $10^{-4}$, which is within the current 68\% confidence interval from measurements by the {\it Planck} satellite. (Notice that allowing mild scale-dependence in $a_1$, $a_2$ or $a_3$ {\it would} generate interesting shifts to the spectral index. The example of scale-dependent $a_1$ was discussed in \cite{Bramante:2013moa}.)

The bispectrum observed in biased sub-volumes is also shifted. For the exact local ansatz, the change in the observed level of non-Gaussianity was discussed in \cite{Nelson:2012sb,LoVerde:2013xka}. When the super cosmic variance shifts to the power spectrum are scale-dependent, the shape of the observed bispectrum changes:
\be
B^{\rm obs}(k_1,k_2,k_3)=P_{\chi}(k_1)P_{\chi}(k_2)\frac{\fnl N_2(k_1,k_2,k_3)}{[1 + \fnl \Phi_1^{\rm SCV}(k_1)] [1 + \fnl \Phi_1^{\rm SCV}(k_2)]}+\mbox{cyc.}\;
\ee
However, as with the scale-dependent corrections to the power spectrum, the changes in shape are too small to be observationally relevant. We will find much more interesting effects from the inclusion of cubic terms in the ansatz for the non-Gaussian field, which we turn to next.

\section{Cubic Terms}
\label{sec:cubic}

In this section we establish a method to generate cubic terms that induce the four quadratic terms of Eq.\eqref{Phi_quad} in biased sub-volumes.     

\subsection{Generation of cubic terms}
\label{sec:cubicRules}

In general, we consider cubic terms that take the form
\be\label{cubic_form}
\partial^{\beta_5}(\partial^{\beta_4}\phi\partial^{\beta_3}(\partial^{\beta_2}\phi\partial^{\beta_1}\phi))\;.
\ee
Similarly to the rules we used to limit the quadratic terms, we consider only cubic terms that satisfy the following restrictions:
\begin{enumerate}
\item{$\sum\beta_i=0$ to maintain scale-invariance,}
\item{$\beta_{1,2,4} \geq 0$ to ensure that the induced quadratic terms do not depend on $\partial^{-1}\phi_l$ or higher inverse derivatives, controlling sensitivity to the infrared (IR) scale.}
\item{$|\beta_i|\leq2$ restricts to a minimal set of terms to consider.} Note that conditions 1-3 imply $\beta_5 \leq 0$.
\item{$\beta_1+\beta_2+\beta_3 \geq 0$ to ensure that the induced linear terms are no more sensitive to the IR scale than in local non-Gaussianity (logarithmic sensitivity).
\item At least one of $\beta_1$, $\beta_2$, or $\beta_4$ must be zero and the $\{\beta_1,\beta_2\}$ pairs should be drawn from the same set of values as the $\{\alpha_1, \alpha_2\}$ pairs in Eq.\eqref{Phi_quad}. This ensures that all the leading order quadratic terms induced in biased sub-volumes are one of those in Eq.\eqref{Phi_quad}.}
\end{enumerate}
As a guide to understanding the origin of these restrictions, an example of the quadratic terms induced in biased sub-volumes by a cubic term is given in Appendix \ref{Append-FT-b13}.
There are 18 cubic terms, listed in Appendix \ref{appen:cubicterms}, which satisfy these five rules. As with the quadratic terms, it is certainly possible to consider a larger set of terms that have interesting behavior in biased sub-volumes. In Section \ref{sec:families} we discuss the consequences of the minimal set of terms and also comment on the role of some terms we have discarded. Appendix \ref{appen:cubicterms} also gives the cubic kernel, $N_{3}(\vv p_1,\vv p_2,\vv p_3,\vv k)$, from these 18 terms. 

The addition of this cubic functional adds a leading order trispectrum to the model:
\bea
\label{eq:fourpoint}
\langle\Phi(\mathbf k_1)\Phi(\mathbf k_2)\Phi(\mathbf k_3)\Phi(\mathbf k_4)\rangle&=&(2\pi)^3\delta^3_D(\mathbf k_1+\mathbf k_2+\mathbf k_3+\mathbf k_4)\;T_{\Phi}(\vv k_1,\vv k_2,\vv k_3,\vv k_4)\nn\\
T_{\Phi}(\vv k_1,\vv k_2,\vv k_3,\vv k_4)&=&g_{\rm NL}P_{\phi}(k_1)P_{\phi}(k_2)P_{\phi}(k_3)N_3(\vv k_1,\vv k_2,\vv k_3,\vv k_4)+\mbox{cyc.}\;
\eea
where there are three additional terms in the $+\mbox{cyc}$. Notice that our definition of $\gnl$ as the amplitude of the trispectrum goes beyond the usual definition. The typical conventions for defining the amplitudes of trispectra in various momentum configurations are \cite{Byrnes:2006vq, Smith:2011if}:
\bea
{\rm Standard\;trispectra\;conventions:}\;\;\;\gnl^{\rm standard}&\equiv&\frac{1}{6}\lim_{k_{1}\rightarrow0}\frac{T_{\Phi}(\vv k_1,\vv k_2,\vv k_3,\vv k_4)}{P_{\Phi}(k_{1})P_{\Phi}(k_2)P_{\Phi}(k_3)}\nn\\
\tau_{\rm NL}&\equiv&\frac{1}{4}\frac{9}{25}\lim_{|\vv k_{1}+\vv k_2|\rightarrow0}\frac{T_{\Phi}(\vv k_1,\vv k_2,\vv k_3,\vv k_4)}{P_{\Phi}(|\vv k_{1}+\vv k_2|)P_{\Phi}(k_1)P_{\Phi}(k_3)}
\eea
Because we have imposed $|\beta_i|\leq2$ (and specifically $\beta_3\geq-2$), all of our trispectra have $\tau_{\rm NL}=0$. Current constraints on the trispectrum from {\it Planck} satellite data are (allowing both standard shapes to be non-zero) $\tau_{\rm NL}=0.3\pm0.9\times10^4$, $\gnl^{\rm standard}=-1.2\pm2.8\times 10^5$ \cite{Feng:2015pva} at $68\%$ CL, or assuming only one non-zero template at a time $\tau_{\rm NL}<2800$ at $95\%$ CL \cite{Ade:2013ydc}, $\gnl^{\rm standard}=-1.3\pm1.8\times 10^5$ at $68\%$ CL \cite{Feng:2015pva}.
\subsection{Constraints on cubic terms from their contributions to the power spectrum}
As with the quadratic terms, we impose additional restrictions based on the behavior of (classical) loop corrections from the cubic terms, requiring that UV divergences are not stronger than the log divergence of the tree-level case. 

The cubic field $\Phi_3$ contributes to the power spectrum at order $g_{\rm NL}^2$ (from $\langle \Phi_3(\mathbf k_1) \Phi_3(\mathbf k_2) \rangle$):
\bea
\delta P_{\rm \Phi_3}(k) = \gnl^2
\int \!\!\frac{d^3p_1}{(2\pi)^3}\frac{d^3p_2}{(2\pi)^3}\,d^3p_3\, N_{3}^{2}(\vv p_1,\vv p_2,\vv p_3,\vv k)P(p_1)P(p_2)P(p_3)\, \delta^{3}(\vv k - \vv p_1 -\vv p_2 - \vv p_3)\;.
\label{1loop_P_13}
\eea 
Since this contribution has two momentum integrals (two classical loops), there are several different momentum configurations to consider. The resulting constraints (details appear in Appendix \ref{Append-2loop-P}) are
\bea \label{twoloop_constraint_final}
b_{10} = b_{14}= b_{16}= b_{18} =  0 \nn \\
b_3+b_4 = 0 \nn \\
b_6 + b_7 = 0 \nn \\
b_{12} + b_{13} = 0 \nn \\ 
b_{11} + b_{15} + b_{17} = 0 \nn \\
2(b_5+b_8) + b_6 + b_9  + b_{12} = 0
\eea
After imposing these restrictions, we are left with a 9 parameter set of cubic terms to explore. Corrections to the bispectrum from the cubic term (proportional to $\fnl\gnl$ at lowest order) do not give any additional constraints that are not covered by conditions imposed on the quadratic term.

\subsection{The field observed in biased sub-volumes}

To work out the statistics observed in biased sub-volumes of a field described by Eq.\eqref{ansatz} and the cubic functional from the previous section, we again split $\Phi(\vv k)$ into long and short wavelength components:
\bea
\Phi_s(\vv k) = \phi_s(\vv k) + \rm f_{NL} \Phi_2^{(s)}(\vv k) + g_{NL} \Phi_3^{(s)}(\vv k) \\
\Phi_l(\vv k) = \phi_l(\vv k) + \rm f_{NL} \Phi_2^{(l)}(\vv k) + g_{NL} \Phi_3^{(l)}(\vv k)
\eea
where
\begin{align} \label{split_s_3}
\Phi_3^{(s)}(\vv k) & = \frac{1}{3!} \iiint \frac{d^3p_1}{(2 \pi)^3} \frac{d^3p_2}{(2 \pi)^3} d^3 p_3 \, N_{3}(\vv p_1,\vv p_2,\vv p_3,\vv k) \, \delta^3(\vv k - \vv p_1 - \vv p_2 - \vv p_3)\, \Theta(|\vv p_1 + \vv p_2 + \vv p_3| - k_*) \nn \\
& \bigg[ \phi(\vv p_1)\phi(\vv p_2)\phi(\vv p_3)-\sum_{\substack{i=1\\k\neq j\neq i}}^3\phi(\vv p_i)\langle\phi(\vv p_j)\phi(\vv p_k)\rangle \bigg] + \rm cyc
\end{align}
\begin{align} \label{split_l_3}
\Phi_3^{(l)}(\vv k) & = \frac{1}{3!} \iiint \frac{d^3p_1}{(2 \pi)^3} \frac{d^3p_2}{(2 \pi)^3} d^3 p_3 \, N_{3}(\vv p_1,\vv p_2,\vv p_3,\vv k) \, \delta^3(\vv k - \vv p_1 - \vv p_2 - \vv p_3)\, \Theta(k_* - |\vv p_1 + \vv p_2 + \vv p_3|) \nn \\
& \bigg[ \phi(\vv p_1)\phi(\vv p_2)\phi(\vv p_3)-\sum_{\substack{i=1\\k\neq j\neq i}}^3\phi(\vv p_i)\langle\phi(\vv p_j)\phi(\vv p_k)\rangle \bigg] + \rm cyc
\end{align}

The step function in Eq.\eqref{split_s_3} allows several different types of terms (analogous to the two terms in Eq.\eqref{eq:shortPhi}). In particular, there is a term with all three fields $\phi_s$ which ensures that all sub-volumes have the same cubic term as the parent volume. There is also a term with a single $\phi_l$ that contributes to the quadratic term in biased sub-volumes. In other words, the quadratic term that describes the non-Gaussian field in the sub-volume (see Eq.(\ref{eq:generalObserved})) is
\be
(\fnl\Phi_2)^{\rm obs}=\fnl(\Phi_2+\Phi_2^{SCV})
\ee
where
\bea
\fnl\Phi_2^{SCV}=\lefteqn{\gnl\Big[\phi_l\bigg\{
(3b_1+b_2+b_3+b_8+b_{17})\phi_s^2
+(b_2+b_{13})\partial^{-1}(\phi_s\partial\phi_s)
}
\nn \\
&& \hspace{2cm}
+(b_3+b_6+2b_{11}+ b_{13} + 2b_{17})\partial^{-2}(\phi_s\partial^2\phi_s)
+(b_4+b_7+b_{12} + 2b_{15})\partial^{-2}(\partial\phi_s)^2
\bigg\}
\nn \\
&& \hspace{1cm}+\partial\phi_l\bigg\{
(b_2+2b_4)\phi_s\partial^{-1}\phi_s
+(b_5+2b_7+b_{15})\partial^{-1}(\phi_s^2)
+b_9\partial^{-1}\partial\phi_s\partial^{-1}\phi_s
\nn \\
&& \hspace{2cm}
+(b_9+2b_{12})\partial^{-2}(\phi_s\partial\phi_s) + b_{13}\partial^{-2}(\partial^{2}\phi_s\partial^{-1}\phi_s)
\bigg\}
\nn \\
&& \hspace{1cm}+\partial^2\phi_l\bigg\{
b_3\phi_s\partial^{-2}\phi_s
+b_6\partial^{-1}(\phi_s\partial^{-1}\phi_s)
+b_{11}\partial^{-2}(\phi_s^2) + b_{13}\partial^{-3}(\phi_s\partial\phi_s)
\bigg\}\Big]\;- {\rm [E.V.]}.
\label{Phi_cubic_quad}
\eea
Here the subscript `SCV' indicates that this is a super cosmic variance contribution to the quadratic term and we have used the conditions from Eq.\eqref{twoloop_constraint_final} to remove some of the original 18 $b_i$. By design, the largest cosmic variance terms (those terms proportional $\phi_l$) regenerate the four quadratic terms which span the local, equilateral, and orthogonal bispectra types. The contributions proportional to $\partial\phi_l$ or $\partial^2\phi_l$ are subleading, and likely to be unobservably small (although we will comment further on those terms in the next section). Focussing on the leading contributions, we can more clearly make contact with our discussion of bispectral shapes (Eq.(\ref{Phi_quad})) by writing 
\be
\Phi_2^{SCV}=\frac{\gnl\phi_l}{\fnl}\Big\{[A_1\phi_s^2+A_2\partial^{-1}(\phi_s\partial\phi_s)+A_3\partial^{-2}(\phi_s\partial^2\phi_s)+A_4\partial^{-2}(\partial\phi_s)^2]+\dots- {\rm [E.V.]}\Big\}
\ee
There is also a term in Eq.\eqref{split_s_3}, $\Phi_3^{(s)}$, with a single $\phi_s$: the cubic terms also contribute to the linear term in biased sub-volumes. Including the shift to the linear term from $\Phi_2$ (see Eq.(\ref{eq:PhiLinQuad})), the induced linear term is
\bea \label{Phi_cubic_linear}
\Phi_1^{\rm SCV}&=&\;\;\fnl\bigg[\phi_s\Big\{
(2a_1+a_2+a_3)\phi_l+(a_2-2a_3)\partial\phi_l\frac{1}{k}+a_3\partial^2\phi_l\frac{1}{k^2}\Big\}\bigg]\nn\\
&&+\gnl\bigg[\phi_s\bigg\{[3b_1+b_2+b_3+b_5+b_6+2b_8+b_{11}+2b_{17}] \phi_l^2+ [b_2+b_9+b_{13}] \partial^{-1}(\phi_l\partial\phi_l)+
\nn \\
&&\;\;\;\;\;\;\;\;\;b_3\partial^{-2}(\phi_l\partial^2\phi_l) +b_4 \partial^{-2}(\partial\phi_l)^2\bigg\}\nn\\
&& + \partial^{-1}\phi_s\bigg\{[b_2+2b_4+2b_5+2b_7+b_9+2b_{12}+2b_{15}](\phi_l\partial\phi_l) + b_6\partial^{-1}(\phi_l\partial^2\phi_l)+b_7 \partial^{-1}(\partial\phi_l)^2+[b_8+b_{15}]\partial(\phi_l^2)\bigg\} \nn\\
&& +\, \partial^{-2}\phi_s\bigg\{[b_3+b_6+2b_{11}+b_{13}](\phi_l\partial^2\phi_l)+[b_9+b_{12}](\partial\phi_l)^2+b_{17}\partial^2(\phi_l)^2\bigg\} \nn\\
&&+\, \partial^{-3}\phi_s\bigg\{b_{13}(\partial^2\phi_l\partial\phi_l)\bigg\} \bigg]- {\rm [E.V.]}
\eea

All the terms inside the first set of curly brackets in the $\gnl$ term generate SCV of the same order (that is, $\langle(\phi_l^2)^2\rangle\sim\langle(\partial^{-1}(\phi_l\partial\phi_l))^2\rangle$, etc), while the terms on the fourth line and below give contributions that are significantly smaller. Furthermore, since the individual terms inside each set of $\gnl$ curly brackets have the same dimension, there is no way to distinguish them. The only relevant feature is the degree of scale-dependence of the prefactor.

Finally, notice that the short-wavelength field contains terms where, for example, $p_1$, $p_2$, $p_3>k_*$ but $|\vv p_1+\vv p_2|<k_*$. This momentum range allows us to consider the limit of correlation functions where momentum modes accessible in a sub-volume sum up to a long wavelength mode. The information in that limit is about the variation of lower order correlations over spatial distances the size of the long wavelength mode \cite{Smith:2011if, Senatore:2012wy}. 


\section{Families of cubic terms from the super cosmic variance point of view}
\label{sec:families}
In this section, we point out various interesting special cases of the general formulae from the previous section. In each of the subsections below we impose additional constraints (beyond the loop correction considerations) that restrict the super cosmic variance contributions to the bispectrum and power spectrum. That is, we impose constraints on the various combinations of the $b_i$ parameters in Equations (\ref{Phi_cubic_quad}) and (\ref{Phi_cubic_linear}).

\subsection{General results}
Before considering the results of the previous section organized by the squeezed limit behavior of the bispectrum, we first note some general features of the possible super cosmic variance consequences from our cubic terms. 
\begin{itemize}
\item We find cubic terms that shift the bispectrum at order $\phi_{\ell}$ but that give no shift to the power spectrum at order $\phi_s$ (the terms in Eq.(\ref{Phi_cubic_linear}) proportional to $\phi_l^2$, $\partial^{-1}(\phi_l\partial\phi_l)$, $\partial^{-2}(\phi_l\partial^2\phi_l)$ and $\partial^{-2}(\partial\phi_l)^2$ individually vanish). This set of cubic terms only generates $\Phi_2^{\rm SCV}$ where the $\{A_i\}$ are linear combinations of $\{1,0,-2,2\}$ and $\{1,-2,0,0\}$. 
\item We find cubic terms that shift the power spectrum at order $\phi_s$ but that generate no significant shift to the bispectrum (all terms proportional to $\phi_{\ell}$ in the induced quadratic term individually vanish). The $\Phi_1^{\rm SCV}$ generated by any terms in this set are indistinguishable at order $\phi_s$. Some can, in principle, be distinguished by the relative amplitude of the sub-leading terms (proportional to $\partial^{-1}\phi_s$ and $\partial^{-2}\phi_s$).
\item We find cubic terms that generate neither a shift proportional to $\phi_{\ell}$ in the induced quadratic term, Eq.(\ref{Phi_cubic_quad}), nor a shift proportional to $\phi_s$ in the induced linear term, Eq.(\ref{Phi_cubic_linear}). From an observational point of view, these cubic terms generate no super cosmic variance of an interesting size. One of these induces no sub-leading shift to the power spectrum and at most a $\partial^2\phi_l$ shift to the bispectrum and so is a candidate for a trispectrum template consistent with single clock inflation. The kernel and trispectra for this case can be found in Appendix \ref{AppendExTrispect}.

\end{itemize}

To summarize, our set of cubic terms demonstrates that it is possible to find examples with significant cosmic variance in the bispectrum but not the power spectrum, in the power spectrum but not the bispectrum. As may be expected, our set also contains terms that generate no interesting cosmic variance. Furthermore, there is, in general, a degeneracy in the induced lower order terms: Multiple cubic terms (eg, terms that give different trispectra) can generate indistinguishable $\Phi_1^{\rm SCV}$, $\Phi_2^{\rm SCV}$. 

\subsection{Cubic terms that induce equilateral type bispectra in biased sub-volumes}
If we impose the conditions $4A_1+2A_2+2A_3=0$, and $2A_2-4A_3=0$, the SCV induced quadratic term will generate a bispectrum of the equilateral shape (the condition $A_3+A_4=0$ is already enforced by the loop constraints on $\Phi_3$). Most cubic terms satisfying those constraints shift the power spectrum at leading order. However, there are cases that generate an equilateral bispectrum but no leading order shift to the power spectrum (no term proportional to $\phi_s$ in Eq.(\ref{Phi_cubic_linear})). This case may, in principle, be distinguished from a purely equilateral bispectrum by a sub-leading term (proportional to $\partial\phi_l$), although in practice that contribution is quite small. There is also a cubic term whose SCV contribution gives an equilateral bispectrum at leading order and no contributions to the bispectrum of size $\partial\phi_l$ (although this solution does affect the power spectrum at order $\phi_s$).

\begin{figure}[htbp]
 \centering
 \includegraphics[scale=0.33]{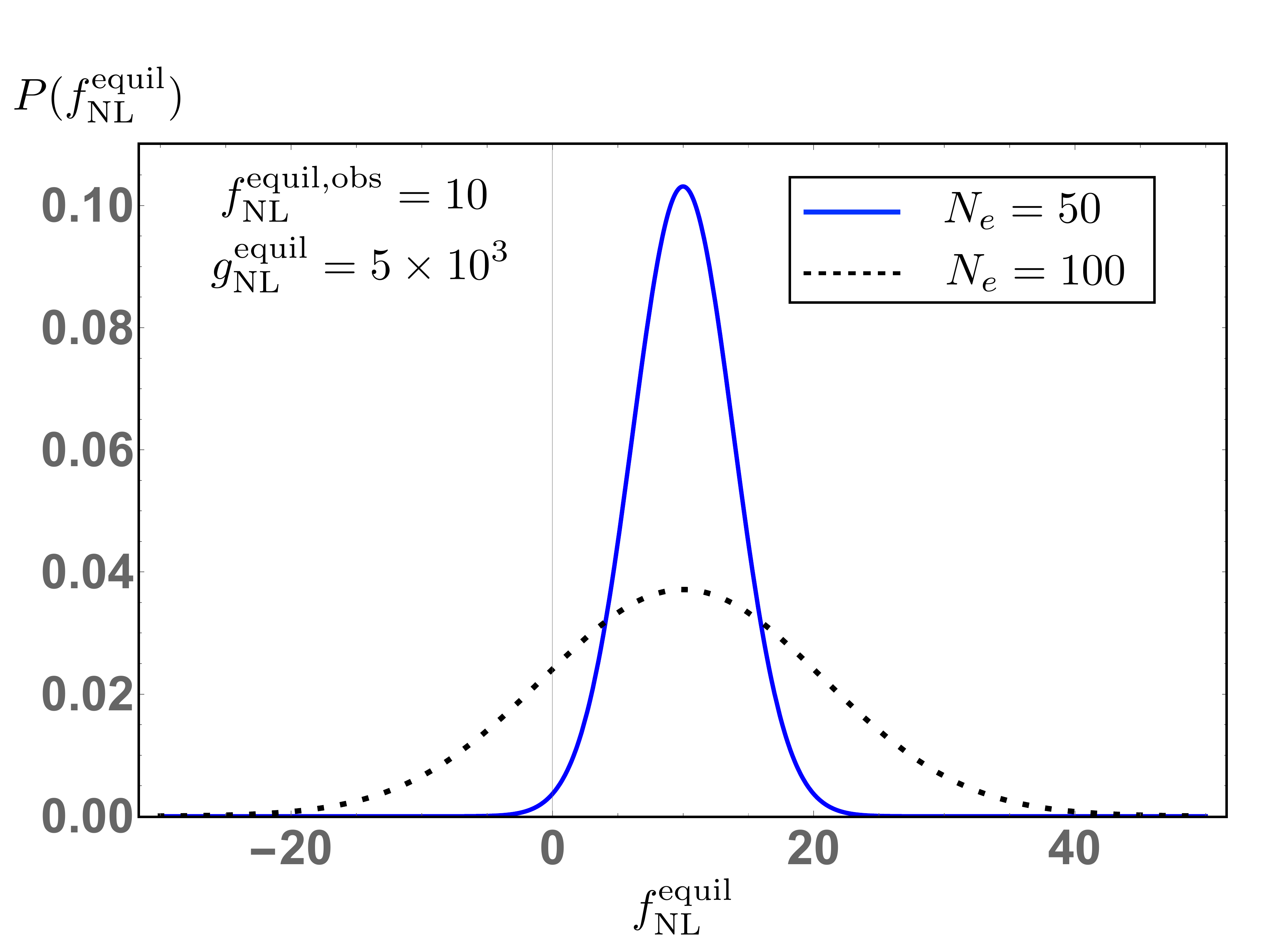}
 \caption{Assuming that a single sub-volume measurement of $\fnl^{\rm equil}=10$, this plot shows the probability distribution for the value of $\fnl^{\rm equil}$ that could be present in the (mean statistics of the) parent volume. Both curves assume the parent volume has a power spectrum consistent with {\it Planck} satellite observations and $\gnl^{\rm equil}=5\times 10^3$ (where $\gnl^{\rm equil}$ is defined as the amplitude of the non-local cubic term specified in Eq.(\ref{eq:scarybi})). If there are no further SCV effects from higher order terms, any sub-volume will have the same power spectrum and $\gnl^{\rm equil}$ as the parent volume. The solid (blue) curve considers a parent volume larger than the sub-volume by a factor of 50 extra e-folds. The dotted (black) curve show a parent volume larger by a factor of 100 extra e-folds. For the best fit {\it Planck} power spectrum and $N_e=50$, $100$, the RMS variance of the long wavelength modes is $\sqrt{\langle\phi_l^2\rangle}\approx0.0008$ and $0.0022$ respectively. }
 \label{fig:shift_fnl_equil}
\end{figure}

We can define an equilateral family of statistics by those $\Phi_n$ whose SCV contributions induce only an equilateral shape bispectrum at leading order and shift the power spectrum by at most terms proportional to $\partial^{-2}\phi_s$ (the same effect that an equilateral bispectrum gives). Note that although this family is distinct from the set of correlations with no SCV at all, it is not observationally distinct if only the power spectrum and the shape of the bispectrum have been measured. A set of non-zero $\{b_i\}$ for the cubic term that induces the equilateral bispectrum but no significant shift to the power spectrum is 
\be
\label{eq:scarybi}
{\rm Equilateral \;family\; cubic\;term:}\;\;\;\;\;b_1=-5/3\;,\;\;\;b_2=2\;,\;\;\;b_5=3\;,\;\;\;b_9=-4\;,\;\;\;b_{12}=-2\;,\;\;\;b_{13}=2.
\ee
The field observed in biased sub-volumes in this scenario (assuming only an equilateral bispectrum in the mean statistics) takes the form
\bea
\Phi^{\rm obs}(\vv k)&=&\phi(\vv k)\\
&&+\frac{(\fnl^{\rm equil}+\gnl^{\rm equil}\phi_l)}{2!(2\pi)^3}\int d^3p_1d^3p_2 \,[\phi(\vv p_1)\phi(\vv p_2)-\langle\phi(\vv p_1)\phi(\vv p_2)\rangle]\,N_2^{\rm equil}(\vv p_1, \vv p_2,\vv k)\delta^{3}(\vv k-\vv p_1-\vv p_2)\nn\\
&&+\frac{g_{\rm NL}^{\rm equil}}{3!(2 \pi)^6}\prod_{\ell=1}^3\int  d^3 p_\ell \left[\phi(\vv p_1)\phi(\vv p_2)\phi(\vv p_3)-\sum_{\substack{i=1\\k\neq j\neq i}}^3\phi(\vv p_i)\langle\phi(\vv p_j)\phi(\vv p_k)\rangle\right]N_3^{\rm equil}(\vv p_1,\vv p_2,\vv p_3,\vv k)\delta^{3}(\vv k-\sum_{\ell=1}^3\vv p_{\ell})\nn
\eea
where $N_2^{\rm equil}$ is Eq.(\ref{N_2}) with $\{a_i\}=\{-3,4,2,-2\}$ (see Eq.(\ref{equil})) and $N_3^{\rm equil}$ is Eq.(\ref{N_3}) with the set of $\{b_i\}$ from Eq.(\ref{eq:scarybi}). Figure \ref{fig:shift_fnl_equil} shows an example quantification of the qualitatively important result for this scenario: there is a range of values of $\fnl^{\rm equil}$ in the mean or parent statistics that are consistent with a particular value of $\fnl^{\rm equil, obs}$ in sub-volumes. Note that the parent volume can have an input $\fnl^{\rm equil}$ of either sign. Appendix \ref{AppendExTrispect} contains more details of some of the interesting cubic terms that induce an equilateral bispectrum in biased sub-volumes.

\subsection{Cubic terms that induce local type bispectra in biased sub-volumes}
If we impose the conditions $A_2=A_3=A_4=0$, the SCV induced quadratic term will generate a local shape bispectrum. All such cubic terms also shift the power spectrum at order $\phi_s$ and most also give sub-leading shifts. If we define the local family as the set $\Phi_n$ whose SCV contributions induce only a local bispectrum and only shift the power spectrum as $\phi_s$ (no sub-leading terms, to match the behavior of the local bispectrum), our ansatz contains several cubic terms in the family. If we are more restrictive and insist that the induced bispectrum has no piece proportional to either $\partial\phi_l$ or $\partial^2\phi_l$ in the bispectrum, the only solution is $b_i=0$ except for $i=1$ (the local $g_{\rm NL}$ term). Super cosmic variance from a local cubic term and beyond was previously considered in \cite{Nelson:2012sb, Byrnes:2006vq}.

\subsection{Cubic terms that induce orthogonal type bispectra in biased sub-volumes}
If we impose the conditions $2A_1+A_2+A_3=0$, the SCV induced quadratic term will generate a bispectrum with a squeezed limit that diverges as $1/k_l^2$ (the condition $A_3+A_4=0$ is already enforced by the loop constraints on $\Phi_3$). All such cubic terms in our set generate no shift to the power spectrum at leading order. Notice that this does not insist on the orthogonal template, only the squeezed limit behavior. However, it is possible to have the exact orthogonal template bispectrum induced with or without also inducing a shift to the power spectrum. As in the local and equilateral case, we can define the orthogonal family as the set of terms in $\Phi[\phi(\vv{x})]=\phi(\vv{x})+ f_{\rm NL} \Phi_2[\phi(\vv{x})]+g_{\rm NL} \Phi_3[\phi(\vv{x})]+\cdots\;$ that has no stronger than $\phi_s/k$ SCV in the power spectrum and maintains a bispectrum with only $1/k_l^2$ divergence. 

\section{Discussion}
\label{sec:conclusions}

In this work we have considered non-Gaussian fields built from non-local real space expressions with up to cubic dependence on a Gaussian field. This ansatz, Eq.(\ref{ansatz}), can generate a variety of tree-level bispectra and trispectra, including shapes consistent with either single-clock or multi-field inflationary models. By examining the ways in which statistics in spatial sub-volumes can differ from the mean statistics of the parent volume we draw the following important conclusions:

\begin{itemize}
\item Terms at order $n$ in a non-local expression for a non-Gaussian field $\Phi(\vv x)$ can generically shift the amplitudes and momentum dependence of the power spectrum and correlations up to order $\langle\Phi(k_1)...\Phi(k_n)\rangle$ in biased sub-volumes, but need not shift {\it all} of them. 
\item More specifically, super cosmic variance effects can induce bispectra in biased subvolumes to match any of the equilateral, orthogonal, or local templates. The amplitude and sign of the induced bispectrum depends in a degenerate way on the background over- or under-density of the sub-volume as well as the amplitude and sign of the mean trispectrum. Depending on the specific form of the trispectrum, the power spectrum may or may not display a dependence on background density within the sub-volume.

\end{itemize}
 
Mathematically, our results are very reasonable: non-Gaussian statistics at each order are independent, so of course measuring the 3-point function alone constrains neither the shape nor the amplitude of higher order correlations. Furthermore, different $n$-point correlations can induce indistinguishable contributions to lower order correlations in biased sub-volumes. When neither the bias nor the higher order correlations can be measured, there is a large degeneracy of correlation functions (and so inflation models) that can be consistent with just a few measurements. Specifically, we have demonstrated that a limited set of measurements of the primordial density correlations (e.g., a detection of $\fnl^{\rm local}=0$ and $\fnl^{\rm equil}\neq0$) could be consistent with single-clock inflation {\it or} a multi-field scenario. Measurement of a purely equilateral bispectrum does not in and of itself imply that there can be no super cosmic variance at work. To rule out super cosmic variance (at least, to rule it out of observational relevance) we must also constrain any correlation between statistics measured within smaller regions of our own Hubble volume and the background density of those regions.

We have hardly covered the space of non-Gaussian statistics: our set of examples was chosen to be a minimal set that allows us to explore the possible implications of cosmic variance from super horizon modes for non-local, scale-independent bispectra\footnote{There are many other terms one might wish to include to study further aspects of this problem. At the level of the quadratic terms, for example, we have not been sufficiently general to cover the other standard template for single-clock inflation (the `orthogonal' template in the appendix of reference \cite{Senatore:2009gt}). To expand the model to include this template would require allowing $\alpha_i<-2$. At cubic order, we eliminated some terms since they did not induce one of our desired quadratic terms at leading order. Keeping those terms would add more parameters that would relax some of the restrictions on the $b_i$ from removing sub-leading terms (and badly behaved loop contributions).}. We see no obstruction to extending our results to find quartic terms in the non-local expansion that could bias, for example, the trispectrum but not the bispectrum. One might complain that super cosmic variance that evades detection in the simplest measurements (e.g., in the power spectrum) but biases higher order terms (e.g., the trispectrum) would require too much of a conspiracy to be realistic. Addressing this discomfort systematically requires a measure on the space of super cosmic variance effects from multi-field inflation models. It would be interesting to find an inflation model that contains trispectra of the sort we have found or to prove that no model can generate them, nor their higher order generalizations. 

\section{Acknowledgment}
\label{ack}
We thank Louis Leblond for helpful discussions. S.S. and E.N. are funded by NSF Award PHY-1417385. S.P. is supported by the Eberly Research Funds of The Pennsylvania State University. The Institute for Gravitation and the Cosmos is supported by the Eberly College of Science and the Office of the Senior Vice President for Research at The Pennsylvania State University. A.K. is supported in part by the NSF grant PHY-1205388, the Eberly research funds of the Pennsylvania State University, and a Frymoyer Fellowship.

\appendix
\section{Cubic terms and the cubic kernel}
\label{appen:cubicterms}

The functional $\Phi_3[\phi(\vv x)]$ built from the 18 terms derived in Section \ref{sec:cubicRules} is\footnote{We note that three terms - $\partial^{-1}(\partial\phi\partial^{-2}(\partial\phi)^2), \ \partial^{-2}(\partial^2\phi\partial^{-2}(\partial\phi)^2)), \ \partial^{-2}(\partial\phi\partial^{-1}(\partial\phi)^2)$ - have been neglected because they do not have nonzero $\beta_1$, $\beta_2$, or $\beta_4$ and thus do not regenerate the original four quadratic terms, but only terms suppressed by $\partial\phi_l$.
Allowing these terms would introduce new parametric freedom among the 18 terms above, allowing some to be nonzero after the imposition of convergence on loop corrections.}

\bea
\lefteqn{\Phi_3(\vv x) = b_1\phi^3 + b_2\phi\partial^{-1}(\phi\partial\phi)
+b_3\phi\partial^{-2}(\phi\partial^2\phi) 
+b_4\phi\partial^{-2}((\partial\phi)^2)
+b_5\partial^{-1}(\phi^2\partial\phi)
+b_6\partial^{-1}(\phi\partial^{-1}(\phi\partial^2\phi))}
\nn
\\
&& \hspace{0.5cm}
+b_7\partial^{-1}(\phi\partial^{-1}((\partial\phi)^2))
+b_8\partial^{-1}(\phi\partial(\phi^2))
+b_9\partial^{-1}(\partial\phi\partial^{-1}(\phi\partial\phi))
+b_{10}\partial^{-1}(\partial\phi\partial^{-2}(\phi\partial^2\phi))
\nn
\\
&& \hspace{0.5cm}
+b_{11}\partial^{-2}(\phi^2\partial^2\phi)
+b_{12}\partial^{-2}(\phi(\partial\phi)^2)
+b_{13}\partial^{-2}(\partial^2\phi\partial^{-1}(\phi\partial\phi))
+b_{14}\partial^{-2}(\partial^2\phi\partial^{-2}(\phi\partial^2\phi))
\nn
\\
&& \hspace{0.5cm}
+b_{15}\partial^{-2}(\partial\phi\partial(\phi^2))
+b_{16}\partial^{-2}(\phi\partial(\phi\partial\phi))
+b_{17}\partial^{-2}(\phi\partial^2(\phi^2))
+b_{18}\partial^{-2}(\partial \phi \partial^{-1}(\phi\partial^2\phi))\;.
\label{Phi_cubic}
\eea

To compute the kernel associated with $\Phi_3(\vv x)$, we need to take the Fourier transform of each term. As an example, consider the Fourier transform of the 13th term:
\bea
\lefteqn{\partial^{-2}(\partial^2\phi\partial^{-1}(\phi\partial\phi))(\vv k)}
\nn\\
&& 
= \int d^3x 
\frac{1}{k^2}\bigg\{\int \frac{d^3p_1}{(2\pi)^3}p_1^2\phi(\vv p_1)e^{i\vv p_1 \cdot \vv x}
\int \frac{d^3q}{(2\pi)^3}\frac{1}{q}
\Big[
\int \frac{d^3p_2}{(2\pi)^3} \int d^3p_3 
\delta^3(\vv q - \vv p_2 -\vv p_3)\phi(\vv p_2)p_3\phi(\vv p_3)
\Big]e^{i\vv q \cdot \vv x}
\bigg\}
e^{-i\vv k \cdot \vv x}
\nn\\
&&
= \int \frac{d^3p_1}{(2\pi)^3}
\int \frac{d^3p_2}{(2\pi)^3}
\int d^3p_3
\int d^3q\; \delta^3(\vv k - \vv p_1 -\vv q)\delta^3(\vv q - \vv p_2 -\vv p_3)  
\frac{p_1^2p_3}{k^2q}
\phi(\vv p_1)\phi(\vv p_2)\phi(\vv p_3)
\eea
We can use a Dirac delta function to integrate over any of the four momenta ($\vv p_1, \vv p_2, \vv p_3$ or $\vv q$) but to make the calculations more transparent we will treat these possibilities symmetrically and write
\bea
\lefteqn{\partial^{-2}(\partial^2\phi\partial^{-1}(\phi\partial\phi))(\vv k) 
= \frac{1}{3!}
\int \frac{d^3p_1}{(2\pi)^3}
\int \frac{d^3p_2}{(2\pi)^3}
\int d^3p_3
\delta^3(\vv k - \vv p_1 -\vv p_2 -\vv p_3) 
\phi(\vv p_1)\phi(\vv p_2)\phi(\vv p_3)}
\nn \\
&& \times
\Biggl[\frac{p_1p_3^2}{k^2|\vv p_1 + \vv p_2|}
+\frac{p_2p_3^2}{k^2|\vv p_1 + \vv p_2|}
+\frac{p_1p_2^2}{k^2|\vv p_1 + \vv p_3|}
+\frac{p_2^2p_3}{k^2|\vv p_1 + \vv p_3|}
+\frac{p_1^2p_2}{k^2|\vv p_2 + \vv p_3|}
+\frac{p_1^2p_3}{k^2|\vv p_2 + \vv p_3|}
\Biggr]\;.
\eea
Now we can read off the contribution to the cubic kernel from the 13th term:
\be
N_3^{\rm 13th}(\vv p_1, \vv p_2, \vv p_3, k) \equiv
\frac{p_1p_3^2}{k^2|\vv p_1 + \vv p_2|}
+\frac{p_2p_3^2}{k^2|\vv p_1 + \vv p_2|}
+\frac{p_1p_2^2}{k^2|\vv p_1 + \vv p_3|}
+\frac{p_2^2p_3}{k^2|\vv p_1 + \vv p_3|}
+\frac{p_1^2p_2}{k^2|\vv p_2 + \vv p_3|}
+\frac{p_1^2p_3}{k^2|\vv p_2 + \vv p_3|} 
\ee

Applying this procedure to each term gives the full cubic kernel $N_3(\vv p_1, \vv p_2, \vv p_3,k)$:
\bea
\label{N_3}
\lefteqn{N_3(\vv p_1, \vv p_2, \vv p_3, k)}
\nn \\
&=&6b_1
+b_{2}\left[\frac{p_1+p_2}{|\vv p_1 + \vv p_2|}
+\frac{p_1+p_3}{|\vv p_1 + \vv p_3|}
+\frac{p_2+p_3}{|\vv p_2 + \vv p_3|}
\right]+b_{3}\left[\frac{p_1^2+p_2^2}{|\vv p_1 + \vv p_2|^2}
+\frac{p_1^2+p_3^2}{|\vv p_1 + \vv p_3|^2}
+\frac{p_2^2+p_3^2}{|\vv p_2 + \vv p_3|^2}
\right]
\nn\\
&&+2b_{4}\left[\frac{p_1p_2}{|\vv p_1 + \vv p_2|^2}
+\frac{p_1p_3}{|\vv p_1 + \vv p_3|^2}
+\frac{p_2p_3}{|\vv p_2 + \vv p_3|^2}
\right]
+2b_{5}\left[\frac{p_1+p_2+p_3}{k}\right]
\nn\\
&&+b_{6}\left[\frac{p_1^2+p_2^2}{k|\vv p_1 + \vv p_2|}
+\frac{p_1^2+p_3^2}{k|\vv p_1 + \vv p_3|}
+\frac{p_2^2+p_3^2}{k|\vv p_2 + \vv p_3|}
\right] 
+2b_{7}\left[\frac{p_1p_2}{k|\vv p_1 + \vv p_2|}
+\frac{p_1p_3}{k|\vv p_1 + \vv p_3|}
+\frac{p_2p_3}{k|\vv p_2 + \vv p_3|}
\right]\nn\\
&&+2b_{8}\left[\frac{|\vv p_1 + \vv p_2|}{k}+\frac{|\vv p_1 + \vv p_3|}{k}+\frac{|\vv p_2 + \vv p_3|}{k}
\right]
+b_{9}\left[\frac{p_3(p_1+p_2)}{k|\vv p_1 + \vv p_2|}
+\frac{p_2(p_1+p_3)}{k|\vv p_1 + \vv p_3|}
+\frac{p_1(p_2+p_3)}{k|\vv p_2 + \vv p_3|}
\right] 
\nn\\
&&+b_{10}\left[\frac{p_3(p_1^2+p_2^2)}{k|\vv p_1 + \vv p_2|^2}
+\frac{p_2(p_1^2+p_3^2)}{k|\vv p_1 + \vv p_3|^2}
+\frac{p_1(p_2^2+p_3^2)}{k|\vv p_2 + \vv p_3|^2}
\right]+2b_{11}\left[\frac{p_1^2+p_2^2+p_3^2}{k^2}\right] 
+2b_{12}\Bigg[\frac{p_1p_2}{k^2}+\frac{p_1p_3}{k^2}+\frac{p_2p_3}{k^2}
\Bigg] 
\nn\\
&& +b_{13}\left[\frac{p_3^2(p_1+p_2)}{k^2|\vv p_1 + \vv p_2|}
+\frac{p_2^2(p_1+p_3)}{k^2|\vv p_1 + \vv p_3|}
+\frac{p_1^2(p_2+p_3)}{k^2|\vv p_2 + \vv p_3|}
\right]  +b_{14}\left[\frac{p_3^2(p_1^2+p_2^2)}{k^2|\vv p_1 + \vv p_2|^2}
+\frac{p_2^2(p_1^2+p_3^2)}{k^2|\vv p_1 + \vv p_3|^2}
+\frac{p_1^2(p_2^2+p_3^2)}{k^2|\vv p_2 + \vv p_3|^2}
\right] 
\nn\\
&& +2b_{15}\Bigg[\frac{p_1|\vv p_2 + \vv p_3|}{k^2}
+\frac{p_2|\vv p_1 + \vv p_3|}{k^2}
+\frac{p_3|\vv p_1 + \vv p_2|}{k^2}
\Bigg]
\nn\\
&&
+b_{16}\left[\frac{(p_1+p_2)|\vv p_1 + \vv p_2|}{k^2}
+\frac{(p_1+p_3)|\vv p_1 + \vv p_3|}{k^2}
+\frac{(p_2+p_3)|\vv p_2 + \vv p_3|}{k^2}
\right]
\nn\\
&&
+2b_{17}\Bigg[\frac{|\vv p_1 + \vv p_2|^2}{k^2}
+\frac{|\vv p_1 + \vv p_3|^2}{k^2}
+\frac{|\vv p_2 + \vv p_3|^2}{k^2}
\Bigg] +b_{18}\Bigg[\frac{(p_1^2+p_2^2)p_3}{k^2|\vv p_1 + \vv p_2|}
+\frac{(p_1^2+p_3^2)p_2}{k^2|\vv p_1 + \vv p_3|}
+\frac{(p_2^2+p_3^2)p_1}{k^2|\vv p_2 + \vv p_3|}
\Bigg] 
\eea

\section{Example of finding the quadratic terms induced by a cubic term in biased sub-volumes}
\label{Append-FT-b13}

We again use the 13th term, $\partial^{-2}(\partial^2\phi\partial^{-1}(\phi\partial\phi))$, to illustrate how to read off the induced quadratic terms from the limit of a cubic term when one momentum is much smaller than the other two. Choosing any one of the three momenta (e.g., $p_3$) as the long wavelength mode $k_l$ and the other two ($p_1, p_2$) as the short wavelength modes $k_s$, and using the fact that the kernel is symmetric in the $p_i$:
\bea
\lefteqn{\partial^{-2}(\partial^2\phi\partial^{-1}(\phi\partial\phi))(\vv k)}
\nn \\
&&\longrightarrow 3\times\frac{1}{3!}
\int_{\Lambda}^{k_*}\!\! \frac{d^3p_3}{(2\pi)^3}\phi(\vv p_3)
\int_{k_*}^{k_{\rm max}}\!\! \frac{d^3p_1}{(2\pi)^3}
\int_{k_*}^{k_{\rm max}} d^3p_2
\delta^3(\vv k - \vv p_1 -\vv p_2) 
\phi(\vv p_1)\phi(\vv p_2)
\nn \\ 
&& \hspace{1cm} \times 
\Biggl\{
\frac{p_1^2}{k^2}+\frac{p_2^2}{k^2}
+p_3\bigg[\frac{p_1^2}{k^2p_2}+\frac{p_2^2}{k^2p_1}\bigg]
+p_3^2\bigg[\frac{p_1}{k^3}+\frac{p_2}{k^3}\bigg]
\Biggr\}
\nn \\
&&  \equiv \Big[
\phi_l \partial^{-2}(\phi_s\partial^2\phi_s)
+\partial \phi_l \partial^{-2}(\partial^2\phi_s\partial^{-1}\phi_s)
+\partial^2 \phi_l \partial^{-3}(\phi_s\partial\phi_s)
\Big]
\eea
where $\Lambda$ is the largest scale in the problem (eg, corresponding to the mode that first exited during inflation) and $k_*$ is the scale that defines the size of the sub-volume. The corresponding procedure in real space is 
\bea
\lefteqn{\partial^{-2}(\partial^2\phi\partial^{-1}(\phi\partial\phi))
\longrightarrow
\Big[
\partial^{-2}(\partial^2\phi_l\partial^{-1}(\phi_s\partial\phi_s))
+\partial^{-2}(\partial^2\phi_s\partial^{-1}(\phi_l\partial\phi_s))
+\partial^{-2}(\partial^2\phi_s\partial^{-1}(\phi_s\partial\phi_l))
\Big]
}
\nn\\
&& \hspace{2.9cm}=\Big[
\partial^2\phi_l\partial^{-2}(\partial^{-1}(\phi_s\partial\phi_s))
+\phi_l\partial^{-2}(\partial^2\phi_s\partial^{-1}(\partial\phi_s))
+\partial\phi_l\partial^{-2}(\partial^2\phi_s\partial^{-1}(\phi_s))
\Big]
\nn\\
&& \hspace{2.9cm}=\Big[
\phi_l \partial^{-2}(\phi_s\partial^2\phi_s)
+\partial \phi_l \partial^{-2}(\partial^2\phi_s\partial^{-1}\phi_s)
+\partial^2 \phi_l \partial^{-3}(\phi_s\partial\phi_s)
\Big]
\qquad \qquad
\eea
Note that to go from the first line to the second, all the long modes
$\phi_l, \partial \phi_l$ and $\partial^2 \phi_l$
are treated as constants.

\section{The contribution to the power spectrum from the cubic field $\Phi_3$
}
\label{Append-2loop-P}

In this Appendix we derive the constraints on the terms in $\Phi_3$ from their contribution to the power spectrum:
\bea
\lefteqn{\Big\langle \Phi_3(\vv k') \Phi_3(\vv k) \Big\rangle 
\equiv \delta P_{\rm \Phi_3}(k)\,\delta^3(\vv{k'}+\vv{k})}
\nn\\
&& = \bigg(\frac{g_{\rm NL}}{3!}\bigg)^{2}\prod_{\ell=1}^3\int\dthreee p\ell  \prod_{m=1}^3\int\dthreee qm N_3(\vv p_1,\vv p_2,\vv p_3,\vv k) N_3(\vv q_1,\vv q_2,\vv q_3,\vv k') \delta^{3}(\vv k'-\sum_{m=1}^3\vv q_{m}) \delta^{3}(\vv k-\sum_{\ell=1}^3\vv p_{\ell}) 
\nn\\
&& \Bigg\langle \bigg[\phi(\vv p_1)\phi(\vv p_2)\phi(\vv p_3)-\sum_{\substack{i=1\\k\neq j\neq i}}^3\phi(\vv p_i)\langle\phi(\vv p_j)\phi(\vv p_k)\rangle\bigg] \bigg[\phi(\vv q_1)\phi(\vv q_2)\phi(\vv q_3)-\sum_{\substack{i'=1\\k'\neq j'\neq i}}^3\phi(\vv q_i')\langle\phi(\vv q_j')\phi(\vv q_k')\rangle\bigg] \Bigg\rangle 
\nn \\
&& = \gnl^2
\int \!\!\frac{d^3p_1}{(2\pi)^3}\frac{d^3p_2}{(2\pi)^3}\,d^3p_3\, N_{3}^{2}(\vv p_1,\vv p_2,\vv p_3,\vv k)P(p_1)P(p_2)P(p_3)\, \delta^{3}(\vv k - \vv p_1 -\vv p_2 - \vv p_3)\;.
\eea
In order to obtain the constraints on the terms in $N_3$ we need to find those that give divergent integrals in the expression above. The divergences are in the UV, where some of the internal momenta are large, and there are several configurations to consider:

For $k \ll |\vv p_1 + \vv p_2|,p_1 ; k \gg p_2$:
\bea
b_{14} = 0 \nn \\
b_{10} + b_{18} = 0 \nn \\
b_3 + b_4 + b_6 + b_7 + 2b_{11} + b_{12} + b_{13} + b_{14} + 2b_{15} + b_{16} + 2b_{17} + b_{18} = 0
\eea
For $k \ll |\vv p_1 + \vv p_2|,p_1,p_2$: 
\bea
b_{12} + b_{13} + b_{16} = 0 \nn \\
b_{11} + b_{14} + b_{15} + b_{17} + b_{18} = 0 \nn \\
\eea
For $k \ll p_1,p_2; k \gg |\vv p_1 + \vv p_2|$:
\bea
b_{16} = 0 \nn \\
(b_{11} + b_{15} + b_{17}) + (b_{12} + b_{13} + b_{16}) + (b_{11} + b_{14} + b_{15} + b_{17} + b_{18}) = 0 \nn \\
b_6 + b_7 + b_{18} = 0 \nn \\
(2b_5 + b_6 + 2b_8 + b_9 + b_{10} + b_{12}) + (b_{12} + b_{13} + b_{16}) = 0
\eea 

We have verified that other loop corrections (e.g., to the bispectrum) do not give additional constraints. 


\section{Examples of special cubic kernels and the corresponding trispectra}
\label{AppendExTrispect}

A cubic kernel that gives no cosmic variance contributions stronger than $\partial^2\phi_l$ to the power spectrum or bispectrum has $\{b_1=1, b_{11}=3,b_{17}=-3\}$. This is a candidate for a single-clock inflation trispectrum:
\bea
{\rm single\;clock\;candidate:}\;\;\;N_3(\vv k_1,\vv k_2,\vv k_3,\vv k_4)=6+6\left[\frac{k_1^2+k_2^2+k_3^2}{k_4^2}\right]-6\left[\frac{|\vv k_1+\vv k_2|^2+|\vv k_1+\vv k_3|^2+|\vv k_2+\vv k_3|^2}{k_4^2}\right]
\eea
Recall that the trispectrum is related to the cubic kernel by 
\be
T_{\Phi}(\vv k_1,\vv k_2,\vv k_3,\vv k_4)=g_{\rm NL}P_{\phi}(k_1)P_{\phi}(k_2)P_{\phi}(k_3)N_3(\vv k_1,\vv k_2,\vv k_3,\vv k_4)+\mbox{cyc.}\;.
\ee
For the kernel above, the trispectrum has no factors of $|\vv k_i+\vv k_j|$ in the denominator and $\lim_{k_{i}\rightarrow0}T_{\Phi}=0$.

A cubic kernel that induces an equilateral bispectrum and a leading order ($\propto\phi_s$) SCV shift to the power spectrum is $\{b_1=-3,b_2=4,b_3=2,b_4=-2\}$. We start writing the non-Gaussian field up to cubic order with implementing our parameter set for the cubic terms
\begin{center}
Induce equilateral bispectrum and shift PS:
\end{center}
\bea 
 N_3(\vv k_1,\vv k_2,\vv k_3,\vv k_4)&=&-18+4\left[\frac{k_1+k_2}{|\vv k_1 + \vv k_2|}
+\frac{k_1+k_3}{|\vv k_1 + \vv k_3|}
+\frac{k_2+k_3}{|\vv k_2 + \vv k_3|}\right] \nn\\
&& +2\left[\frac{k_1^2+k_2^2}{|\vv k_1 + \vv k_2|^2}+\frac{k_1^2+k_3^2}{|\vv k_1 + \vv k_3|^2}
+\frac{k_2^2+k_3^2}{|\vv k_2 + \vv k_3|^2}\right] \nn\\
&& -4\left[\frac{k_1k_2}{|\vv k_1 + \vv k_2|^2}
+\frac{k_1k_3}{|\vv k_1 + \vv k_3|^2}
+\frac{k_2k_3}{|\vv k_2 + \vv k_3|^2}
\right]
\eea
One can see directly from the trispectrum that this term induces the equilateral bispectrum:
\bea
k_4^3\lim_{k_4\rightarrow0}T_{\Phi}(\vv k_1,\vv k_2,\vv k_3,\vv k_4)&=&P(k_1)P(k_2)\left[-6+4\frac{k_1+k_2}{k_3}+2\frac{k_1^2+k_2^2}{k_3^2}-4\frac{k_1k_2}{k_3^2}\right] + {\rm perm.}
\label{eq:induce_equil}
\eea

The cubic term with $\{b_1=-5/3,b_2=2,b_5=3,b_9=-4,b_{12}=-2, b_{13}=2\}$ also induces the equilateral bispectrum but does not shift the power spectrum at leading order ($\propto \phi_s$). This kernel is
\bea
N_3(\vv k_1,\vv k_2,\vv k_3,\vv k_4)&=&-10
+2\left[\frac{k_1+k_2}{|\vv k_1 + \vv k_2|}
+\frac{k_1+k_3}{|\vv k_1 + \vv k_3|}
+\frac{k_2+k_3}{|\vv k_2 + \vv k_3|}
\right] +6\left[\frac{k_1+k_2+k_3}{k_4}\right]\nn\\
&&-4\left[\frac{k_3(k_1+k_2)}{k_4|\vv k_1+\vv k_2|}+\frac{k_2(k_1+k_3)}{k_4|\vv k_1+\vv k_3|}+\frac{k_1(k_2+k_3)}{k_4|\vv k_2+\vv k_3|}\right]\nn\\
&&-4\left[\frac{k_1k_2+k_1k_3+k_2k_3}{k_4^2}\right]+2\left[\frac{k_3^2(k_1+k_2)}{k_4^2|\vv k_1+\vv k_2|}+\frac{k_2^2(k_1+k_3)}{k_4^2|\vv k_1+\vv k_3|}+\frac{k_1^2(k_2+k_3)}{k_4^2|\vv k_2+\vv k_3|}\right]
\eea

In this case the trispectrum again satisfies Eq.(\ref{eq:induce_equil}) but in addition
\be
{\rm No\;shift\;to\;power\;spectrum:} \;\;k_1^3k_2^3\lim_{k_1,k_2\rightarrow0}T_{\Phi}(\vv k_1,\vv k_2,\vv k_3,\vv k_4)=0\;.
\ee

In the above examples, we have not normalized the coefficients $b_i$. However, for some template definitions it may be useful to require $\sum_i b_i=1$, which can be easily done.

\end{document}